 \documentclass[12pt,a4paper]{article}
 \usepackage{epsf}
\begin{document}
\title{Lepton-Flavor Violating Processes $l_{i}\rightarrow l_{j}\gamma$
in Topcolor Assisted Technicolor Models}
\author{Chongxing Yue$^{(a,b)}$, Qingjun Xu$^{b}$, Guoli Liu$^{b}$
 \\ {\small a: CCAST (World
 Laboratory) P.O. BOX 8730. B.J. 100080 P.R. China}\\
 {\small b: College of Physics and Information Engineering,}\\
 \small{Henan Normal University, Xinxiang  453002. P.R.China}
\thanks{This work is supported by the National Natural Science
 Foundation of China(I9905004), the Excellent Youth Foundation of
  Henan Scientific Committee(9911); and Foundation of Henan Educational
  Committee.}
\thanks{E-mail:cxyue@public.xxptt.ha.cn} }
\date{\today}
\maketitle
\begin{abstract}
\hspace{5mm}We consider the lepton-flavor violating(LFV) processes
$l_{i}\rightarrow l_{j}\gamma$ in the framework of topcolor
assisted technicolor(TC2) models. We find that the new gauge boson
$Z^{\prime}$ predicted by TC2 models can give significantly
contributions to these processes via the flavor changing
couplings. The present experimental bound on the LFV process $\mu
\rightarrow e\gamma$ gives severe constraints on the TC2 models.
In the case that the $Z^{\prime}$ mass $M_{Z}$ is consistent with
other experimental constraints, we obtain constraints on the
lepton mixing factors $K_{\tau\mu}$ and $K_{\tau e}$. The future
LFV experiments will be probe of the TC2 models.
\end {abstract}

\vspace{1.0cm} \noindent {\bf PACS number(s)}: 13.35.-r,12.60.Nz

\newpage
\section{Introduction}
\hspace{5mm}It is well known that the baryon and lepton numbers
are automatically conserved and the tree level flavor changing
neutral currents(FCNC's) are absent in the standard model(SM). The
production cross section of the FCNC process is very small at
one-loop level due to the unitary of CKM matrix. Thus, the FCNC
processes can provide an important test for any new physics beyond
the SM. Any observation of the flavor changing couplings deviated
from that in the SM would unambiguously signal the presence of new
physics.

  The observation of neutrino oscillations\cite{x1,x2} implies that
the individual lepton numbers $L_{e,\mu,\tau}$ are violated,
suggesting the appearance of the lepton-flavor violating(LFV)
processes, such as $\tau \rightarrow \mu\gamma$, $\tau \rightarrow
e \gamma$ and $\mu \rightarrow e \gamma$. However, the branching
ratios of these processes are extremely small in the SM with
right-handed neutrinos. The present experimental limits\cite{x3}
are:
\begin{equation}
B_{r}(\tau \rightarrow \mu\gamma)<1.1\times10^{-6},\hspace{5mm}
B_{r}(\tau\rightarrow e\gamma)<2.7\times10^{-6},\hspace{5mm}
B_{r}(\mu \rightarrow e\gamma)<1.2\times10^{-11}.
\end{equation}
Thus, the observation of any rate for one of these processes would
be a signal of new physics. On very general grounds, theories of
electroweak symmetry breaking(EWSB) often predict LFV effects
within reach of the upcoming experiments\cite{x4}. This fact has
led a lot of theoretical activity involving LFV processes within
some specific models beyond the SM. For example, studies of LFV
processes in supersymmetric(SUSY) models with a gauge unification
group and SUSY models with ``see-saw" neutrinos\cite{x5}, SUSY
models with R-parity violation\cite{x6}, in models with extra
dimensions\cite{x7} and in the Zee model\cite{x8}.

The top quark, with a mass of the order of the weak scale, is
singled out to play a key role in the dynamics of EWSB and flavor
symmetry breaking. There may be a common origin for EWSB and top
quark mass generation. Much theoretical work has been carried out
in connection to the top quark and EWSB. The topcolor assisted
technicolor(TC2) models\cite{x9}, the top see-saw models\cite{x10}
and the flavor universal coloron models\cite{x11} are three of
such examples. These kinds of models generally predict the
existence of colored gauge bosons (top-gluons, colorons),
color-singlet gauge bosons ($Z^{\prime}$) and Pseudo Goldstone
bosons. These new particles are most directly related to EWSB.
Thus, studying the effects of these new particles in various
processes would provide crucial information for EWSB and fermion
flavor physics as well.

 For TC2 models, the underlying interactions, topcolor interactions,
  are non-universal and therefore do not possess a GIM mechanism. When
the non-universal interactions are written in the mass
eigen-basis, it may lead to the flavor changing vertices of the
new gauge bosons, such as $Z^{\prime}\tau e$, $Z^{\prime}\mu e$
and $Z^{\prime}\mu\tau$. Thus the new gauge boson $Z^{\prime}$
might have significant contributions to some LFV
processes\cite{x12,x13}. In this paper, we reexamine the
contributions of the new gauge boson $Z^{\prime}$ to the LFV
processes $l_{i}\rightarrow l_{j}\gamma$($\tau \rightarrow
\mu\gamma$, $\tau \rightarrow e \gamma$ and $\mu \rightarrow e
\gamma$) in the framework of TC2 models. We find that the present
experimental bound on the LFV process $\mu \rightarrow e \gamma$
gives stringent constraints on the lepton mixing factors
$K_{\tau\mu}$ and $K_{\tau e}$. For $M_{Z}=3TeV$,
$K_{\tau\mu}=0.3$, there must be $K_{\tau e}<0.07$ and for
$M_{Z}=3TeV$, $K_{\tau e}=0.3$, there must be $K_{\tau \mu}<0.07$.

  The paper is organized as follows. In section 2 we give the
widths and branching ratios of the LFV processes $l_{i}\rightarrow
l_{j}\gamma$, which arise from the new gauge boson $Z^{\prime}$.
In section 3 we analyse the constraints on TC2 models from the
current experimental bounds on the LFV processes $l_{i}\rightarrow
l_{j}\gamma$ and compare them with that of the LFV process $\mu
\rightarrow 3 e$. Our conclusions are given in section 4.

 \section{The contributions of the new gauge boson $Z^{\prime}$ to
 the LFV  processes $l_{i}\rightarrow l_{j}\gamma$}

 \hspace{5mm}In TC2 models, the ETC interactions have contributions
 to all quark and lepton masses, while the mass of the top quark is
 mainly generated by the topcolor interactions, and EWSB is driven
 by technicolor or a Higgs sector. To maintain electroweak symmetry
 between top and bottom quarks and yet not generate $m_b \simeq
 m_t$, the topcolor gauge group is usually taken to be a
 strongly coupled $SU(3)\otimes U(1)$. The $U(1)$ provides the
 difference that causes only top quarks to condense. At the
 $\Lambda\sim 1 TeV$, the dynamics of a general TC2 model involves
 the following structure \cite{x9,x14}:
 \begin{equation}
   SU(3)_1  \otimes SU(3)_2 \otimes U(1)_{y_1} \otimes U(1)_{y_2}
 \otimes SU(2)_L \longrightarrow SU(3)_{QCD} \otimes U(1)_{EM},
 \end{equation}
 where $SU(3)_1\otimes U(1)_{y_1}$ ($SU(3)_2 \otimes U(1){y_2}$)
 generally couples preferentially to the third (first and second )
 generations. The $U(1)_{y_i}$ are just strongly rescaled versions
 of electroweak $U(1)_y$. This breaking scenario gives rise to the
 topcolor gauge bosons including the color-octet coloron
 $B_{\mu}^A$ and color-singlet extra $U(1)$ gauge boson
 $Z^\prime$.

   The flavor-diagonal couplings of the new gauge boson $Z^\prime$
 to leptons, which are related to the LFV processes
 $l_{i}\rightarrow l_{j}\gamma$, can be written as:
\begin{eqnarray}
\nonumber
\pounds_{Z^\prime}^{FD}&=&-\frac{1}{2}g_{1}\cot\theta^{\prime}
Z_{\mu}^\prime(\bar{\tau}_{L}\gamma^{\mu}\tau_{L}+2
\bar{\tau}_{R}\gamma^{\mu}\tau_{R})
\\
 & & +\frac{1}{2}g_{1}\tan\theta^{\prime}Z_{\mu}^\prime(\bar{\mu}_{L}
 \gamma^{\mu}\mu_{L}+2\bar{\mu}_{R}\gamma^{\mu}\mu_{R}+\bar{e}_{L}
 \gamma^{\mu}e_{L}+2\bar{e}_{R}\gamma^{\mu}e_{R}),
\end{eqnarray}
where $g_{1}$ is the $U(1)_{y}$ coupling constant at the scale
$\Lambda_{TC}$ and $\theta^{\prime}$ is the mixing angle with
$\tan\theta^{\prime}=g_{1}/(2\sqrt{\pi K_{1}})$. For TC2 models,
when the non-universal interactions, topcolor interactions, are
written in the mass eigen-basis, it results in the flavor changing
vertices of the gauge boson $Z^{\prime}$. The flavor changing
couplings of $Z^{\prime}$ to leptons can be written as:
\begin{eqnarray}
 \nonumber \pounds_{Z^\prime}^{FC}&=&-\frac{1}{2}g_{1}Z_{\mu}^\prime
 [K_{\tau\mu}(\bar{\tau}_{L}\gamma^{\mu}\mu_{L}+
2\bar{\tau}_{R}\gamma^{\mu}\mu_{R})  \\
 & & +K_{\tau e}(\bar{\tau}_{L}\gamma^{\mu}e_{L}+
2\bar{\tau}_{R}\gamma^{\mu}e_{R})+K_{\mu e
}\tan^{2}\theta^{\prime}(\bar{\mu}_{L}\gamma^{\mu}e_{L}+
2\bar{\mu}_{R}\gamma^{\mu}e_{R})],
\end{eqnarray}
where $K_{ij}^{'s}$ are the flavor mixing factors.

  Using Eq.3 and Eq.4, we can calculate the contributions of the
gauge boson $Z^{\prime}$ to the muon anomalous magnetic moment
$a_{\mu}$. The result has been given in Ref.[15], which is that,
as long as the $Z^{\prime}$ mass $M_{Z}\geq1TeV$, TC2 models could
explain the observed BNL results of $a_{\mu}$ for $1.1TeV \leq
M_{x_{\mu}} \leq 2.2TeV$. Now, we examine the LFV processes
$l_{i}\rightarrow l_{j}\gamma$($\tau \rightarrow \mu\gamma$, $\tau
\rightarrow e \gamma$, $\mu \rightarrow e \gamma$) which can be
induced from the couplings $Z^{\prime}l_{i}l_{j}$. A general form
of the matrix element, which relates to the LFV processes
$l_{i}\rightarrow l_{j}\gamma$, can be written as:
\begin{equation}
M=e\overline{u}(p_{2})\sigma_{\mu\nu}q^{\nu}(A_{L}p_{L}+A_{R}p_{R})
u(p_{1})\epsilon(q)^{*\mu}.
\end{equation}
Using the formula given by Ref.[16] and Eq.3, Eq.4, the partial
widths can be calculated:
\begin{equation}
\Gamma(\tau\rightarrow\mu\gamma)=\frac{\alpha^{2}m_{\tau}^{5}}{1152
\pi^{2}M_{Z}^{4}C_{W}^{2}}K_{1}K_{\tau\mu}^{2},
\end{equation}
\begin{equation}
\Gamma(\tau\rightarrow e
\gamma)=\frac{\alpha^{2}m_{\tau}^{5}}{1152\pi^{2}M_{Z}^{4}C_{W}^{2}}
K_{1}K_{\tau e }^{2},
\end{equation}
\begin{equation}
\Gamma(\mu\rightarrow
e\gamma)=\frac{\alpha^{3}m_{\mu}^{3}m_{\tau}^{2}}
{512\pi^{2}M_{Z}^{4}C_{W}^{4}}K_{\tau\mu}^{2}K_{\tau e }^{2},
\end{equation}
where $\alpha$ is the electromagnetic coupling constant, $G_{F}$
is the Fermi constant and $S_{W}=\sin\theta_{W}$ which
$\theta_{W}$ is the Weinberg angle. For TC2 models, to obtain the
top quark direction for condensation, we must have
$\cot\theta^{\prime}\gg 1$. In above equations, we have ignored
the high order terms which are proportional to
$(\tan\theta^{\prime})^{2}$ or $(\tan\theta^{\prime})^{4}$ and
approximately take $\theta^{\prime}\approx \theta_W $.

The $e\bar{\nu_{e}}\nu_{\mu}$ is the dominant decay mode of the
lepton $\mu$. If we assume that the total decay width is dominated
by the decay channel $\mu\rightarrow e\bar{\nu_{e}}\nu_{\mu}$,
then we have:
\begin{equation}
B_{r}(\mu \rightarrow e\gamma)=\frac{\Gamma(\mu\rightarrow e
\gamma)}{\Gamma(\mu\rightarrow
e\bar{\nu_{e}}\nu_{\mu})}=\frac{3\pi\alpha^{3}}
{8G_{F}^{2}C_{W}^{4}M_{Z}^{4}}(\frac{m_{\tau}}{m_{\mu}})^{2}
K_{\tau\mu}^{2}K_{\tau e }^{2}.
\end{equation}
The $e\bar{\nu_{e}}\nu_{\mu}$ is one of the dominant decay modes
of the lepton $\tau$. The branching ratio $B_{r}(\tau \rightarrow
e\bar{\nu_{e}}\nu_{\tau})$ has been precisely measured, i.e.
$B_{r}(\tau \rightarrow e\bar{\nu_{e}}\nu_{\tau})=
(17.83\pm0.06)\%$ \cite{x3}. Thus, we can use the branching ratio
$B_{r}(\tau \rightarrow e\bar{\nu_{e}}\nu_{\tau})$ represent the
branching ratios of the LFV processes $\tau \rightarrow \mu\gamma$
and $\tau \rightarrow e\gamma$.
\begin{equation}
B_{r}(\tau \rightarrow \mu\gamma)=\frac{\Gamma(\tau\rightarrow \mu
\gamma)}{\Gamma(\tau\rightarrow e\bar{\nu_{e}}\nu_{\tau})}
\frac{\Gamma(\tau\rightarrow
e\bar{\nu}_{e}\nu_{\tau})}{\Gamma(\tau\rightarrow all)}
=\frac{\pi\alpha^{2}B_{r}(\tau \rightarrow
e\bar{\nu_{e}}\nu_{\tau})}{6G_{F}^{2}C_{W}^{2}M_{Z}^{4}}
K_{1}K_{\tau\mu}^{2},
\end{equation}
\begin{equation}
B_{r}(\tau \rightarrow e\gamma)=\frac{\pi\alpha^{2}B_{r}(\tau
\rightarrow
e\bar{\nu_{e}}\nu_{\tau})}{6G_{F}^{2}C_{W}^{2}M_{Z}^{4}}K_{1}K_{\tau
e }^{2}.
\end{equation}

In the next section, we will use these formula given the
constraints on the TC2 models from the current experimental bounds
on the LFV processes $l_{i}\rightarrow l_{j}\gamma$.

\section{Constraints on the TC2 models}
\hspace{5mm}The new strong interactions may exist at relatively
low scales and may play an integral part in either EWSB or fermion
mass generation. Thus, it is interesting to study current
experimental bounds on the mass of the corresponding gauge bosons.
Ref.[17] gives the limits on the mass of the new gauge boson
$Z^{\prime}$ via studying its corrections to the precisely
measured electroweak quantities at LEP and its effects on bijet
production and single top quark production at Tevatron. To see
whether the precisely measured value of the branching ratio $B_{r}
 (\tau \rightarrow \mu\gamma)$ can give a bound on the $Z^{\prime}$
  mass $M_{Z}$, we give the contour line of $B_{r}(\tau \rightarrow
\mu\gamma)=1.1\times10^{-6}$ in the ($K_{1}$, $M_{Z}$) plane for
$0.2\leq K_{1}\leq1$ and three values of $K_{\tau\mu}$ in Fig.1.
From Fig.1 we can see that the bound on the $Z^{\prime}$ mass
$M_{Z}$ from the experimental value of $B_{r}(\tau \rightarrow
\mu\gamma)$ is very weak. Even if we take the maximum values of
the parameters, i.e. $(K_{1})_{\max}=1$\cite{x9} and
$(K_{\tau\mu})_{\max}=\frac{1}{\sqrt{2}}$\cite{x5}, we only have
$M_{Z}>397GeV$. From Eq.11, we can see that this conclusion is
also applied to the LFV process $\tau \rightarrow e\gamma$. If we
assume $K_{\tau e}\approx K_{\tau\mu}=\frac{1}{\sqrt{2}}$ and take
$K_{1}=1$, $M_{Z}=2TeV$, then we have $B_{r}(\tau \rightarrow
\mu\gamma)=B_{r}(\tau \rightarrow e\gamma)\approx 1.7
\times10^{-9}$, which are far below the present experimental upper
bounds ($\sim10^{-6}$)\cite{x3}. Thus, the present experimental
upper bounds on the LFV processes $\tau \rightarrow \mu\gamma$ and
$\tau \rightarrow e\gamma$ can not give significantly bounds on
TC2 models.

 From Eq.9 we can see that the values of the branching ratio
$B_{r}(\mu \rightarrow e\gamma)$ given by the new gauge boson
$Z^{\prime}$ depend on the $Z^{\prime}$ mass $M_{Z}$ and the
mixing factors $K_{\tau\mu}$ and $K_{\tau e}$. With reasonable
values of $M_{Z}$ ($M_{Z}$ is consistent with other experimental
constraints, such as LEP and Tevatron experiments\cite{x17}), the
current experimental upper bound on the branching ratio $B_{r}(\mu
\rightarrow e\gamma)$ can give constraints on the lepton mixing
factors $K_{\tau\mu}$ and $K_{\tau e}$, which are symmetric on
both factors. In Fig.2, we present our numerical results, which
shows upper bounds on $K_{\tau\mu}$ and $K_{\tau e}$ from
$B_{r}^{exp}<1.2\times10^{-11}$, for $M_{Z}=2TeV$, $3 TeV$ and $ 5
TeV $. From Fig.2 we can see that the present experimental upper
limit on the $B_{r}(\mu \rightarrow e\gamma)$ can put severe
constraints on $K_{\tau\mu}$ and $K_{\tau e}$. If we take
$K_{\tau\mu}=0.3$, $M_{Z}=3TeV$, there must be $K_{\tau e}<0.07$,
and for $K_{\tau e}=0.3$, $M_{Z}=3TeV$, there must be $K_{\tau
\mu}<0.07$.

Comparing the contribution of $Z^{\prime}$ to the LFV process
$\mu\rightarrow 3e$ to that of muon decay process $\mu\rightarrow
e\bar{\nu_{e}}\nu_{\tau}$, which proceeds via the electroweak
gauge boson $W$ exchange, we give the branching ratio
$B_{r}(\mu\rightarrow 3e)$ arising from the $Z^{\prime}$
exchange\cite{x13}. Using $ K_1 \approx \frac{g^2_1
\tan^2\theta'}{4\pi}$ and $ M_W^2= \frac{\sqrt 2 g^{2}_{2}}{8
G_{F}}$, Eq.10 of Ref.[13] can be rewritten as:
\begin{equation}
B_{r}(\mu\rightarrow 3e)=\frac{\Gamma(\mu\rightarrow
3e)}{\Gamma(\mu\rightarrow e\bar{\nu_{e}}\nu_{\tau})}=
\frac{25\pi^{2}
\alpha^{2}\tan^{4}\theta_{W}}{128G^{2}_{F}C_{W}^{4}
M_{Z}^{4}}K_{\mu e}^{2}.
\end{equation}
In Fig.3 we give the contour lines of $B_{r}(\mu \rightarrow
e\gamma)=1.2\times10^{-11}$ and $B_{r}(\mu \rightarrow
3e)=1.2\times10^{-12}$ in the ($K$, $M_{Z}$) plane for
$K_{\tau\mu}\approx K_{\tau e}=K$, which is in the range of
$0.02\leq K\leq0.4$. From Fig.3, we can see that the upper
experimental limits of $B_{r}(\mu \rightarrow e\gamma)$ and
$B_{r}(\mu \rightarrow 3e)$ demand that $M_{Z}$ is larger than
$0.41TeV$ for $K_{\tau\mu}\approx K_{\tau e}=K>0.02$ and $1.42TeV$
for $K_{\mu e}>0.02$, respectively. For $K_{\mu e}= \lambda =
0.22$, $M_{Z}$ must be larger than $4.7TeV$, which is consistent
with the conclusion of Ref.[18]. This conclusion is independent of
the parameter $K_{1}$. For $K_{\tau\mu}= K_{\tau e}\leq 0.25$ and
$K_{\mu e}\leq 0.25 $, the constraints from the precision
experimental value of $B_{r}(\mu \rightarrow 3e)$ on TC2 models
are stronger than that of $B_{r}(\mu \rightarrow e\gamma)$.

Similarly, we can calculate the contributions of the new gauge
boson $Z^{\prime}$ to the LFV processes $\tau\rightarrow
l_{i}l_{j}l_{k}$($\tau\rightarrow 3e$, $\tau\rightarrow ee\mu$,
$\tau\rightarrow e\mu\mu$ and $\tau\rightarrow 3\mu$). Using the
present experimental upper bounds on the branching ratios of these
processes, we can obtain upper bounds on the mixing factor
$K_{\tau e}$, $K_{\tau\mu}$ or $K_{\mu e}$. However, the present
experimental upper bounds on the branching ratios
$B_{r}(\tau\rightarrow l_{i}l_{j}l_{k})$ are of order $10^{-6}$,
which are weaker than those of the LFV processes $\mu\rightarrow
3e$ and $\mu\rightarrow e\gamma$. Thus, compared to the LFV
processes $\mu\rightarrow 3e$ and $\mu\rightarrow e\gamma$, these
processes can not give stringent bounds on TC2 models.

\section{Conclusions}
\hspace{5mm}TC2 theory is an attractive scheme in which there is
an explicit dynamical mechanism for breaking electroweak symmetry
and generating the fermion masses including the heavy top quark
mass. This kind of models predict the flavor changing coupling
vertices of the new gauge bosons, such as $Z^{\prime}\mu e$,
$Z^{\prime}\tau\mu$ and $Z^{\prime}\tau e$. In this paper, we
calculate the effects of these couplings on the LFV processes. We
find that these virtual effects on the LFV processes $\tau
\rightarrow e\gamma$, $\tau \rightarrow \mu\gamma$ and
$\tau\rightarrow l_{i}l_{j}l_{k}$ are far below the present
experimental upper limits on these processes. However, the present
experimental upper bound on the branching ratio $B_{r}(\mu
\rightarrow e\gamma)$ gives severe constraints on TC2 models. In
the case that the $Z^{\prime}$ mass $M_{Z}$ is consistent with
other experimental constraints, we obtain stringent constraints on
the lepton mixing factors $K_{\tau\mu}$ and $K_{\tau e}$. However,
if we assume $K_{\tau\mu}= K_{\tau e}\leq 0.25$ and $K_{\mu e}\leq
0.25 $, we find that the LFV process $\mu \rightarrow 3e$ can give
more severe bound on $M_{Z}$ than that of the LFV process $\mu
\rightarrow e\gamma$. The constraints from $\mu \rightarrow 3e$ on
TC2 models are stronger than that of $\mu \rightarrow e\gamma$. In
the near future, the LFV experiments will increase the sensitivity
to the LFV processes by four or more orders of magnitude. Thus,
the future LFV experiments will be probes of TC2 models.

\newpage
\vskip 2.0cm
\begin{center}
{\bf Figure captions}
\end{center}
\begin{description}
\item[Fig.1:]The contour line of $B_{r}(\tau \rightarrow
\mu\gamma)=1.1\times10^{-6}$ in the ($K_{1}$, $M_{Z}$) plane for
$K_{\tau\mu}=0.707$(solid line), 0.25(dotted-dashed line) and
0.1(dashed line).
\item[Fig.2:]The contour line of $B_{r}(\mu \rightarrow
e\gamma)=1.2\times10^{-11}$ in the ($K_{\tau\mu}$, $K_{\tau e}$)
plane for $M_{Z}=5TeV$ (solid line), $3TeV$(dotted line) and
$2TeV$(dotted-dashed line).
\item[Fig.3:]The contour lines of $B_{r}(\mu \rightarrow
e\gamma)=1.2\times10^{-11}$ and $B_{r}(\mu \rightarrow
3e)=1.2\times10^{-12}$ in the ($K$, $M_{Z}$) plane for
$K_{\tau\mu}\approx K_{\tau e}=K$.
\end{description}

\newpage
\begin{figure}[pt]
\begin{center}
\begin{picture}(250,200)(0,0)
\put(-50,0){\epsfxsize120mm\epsfbox{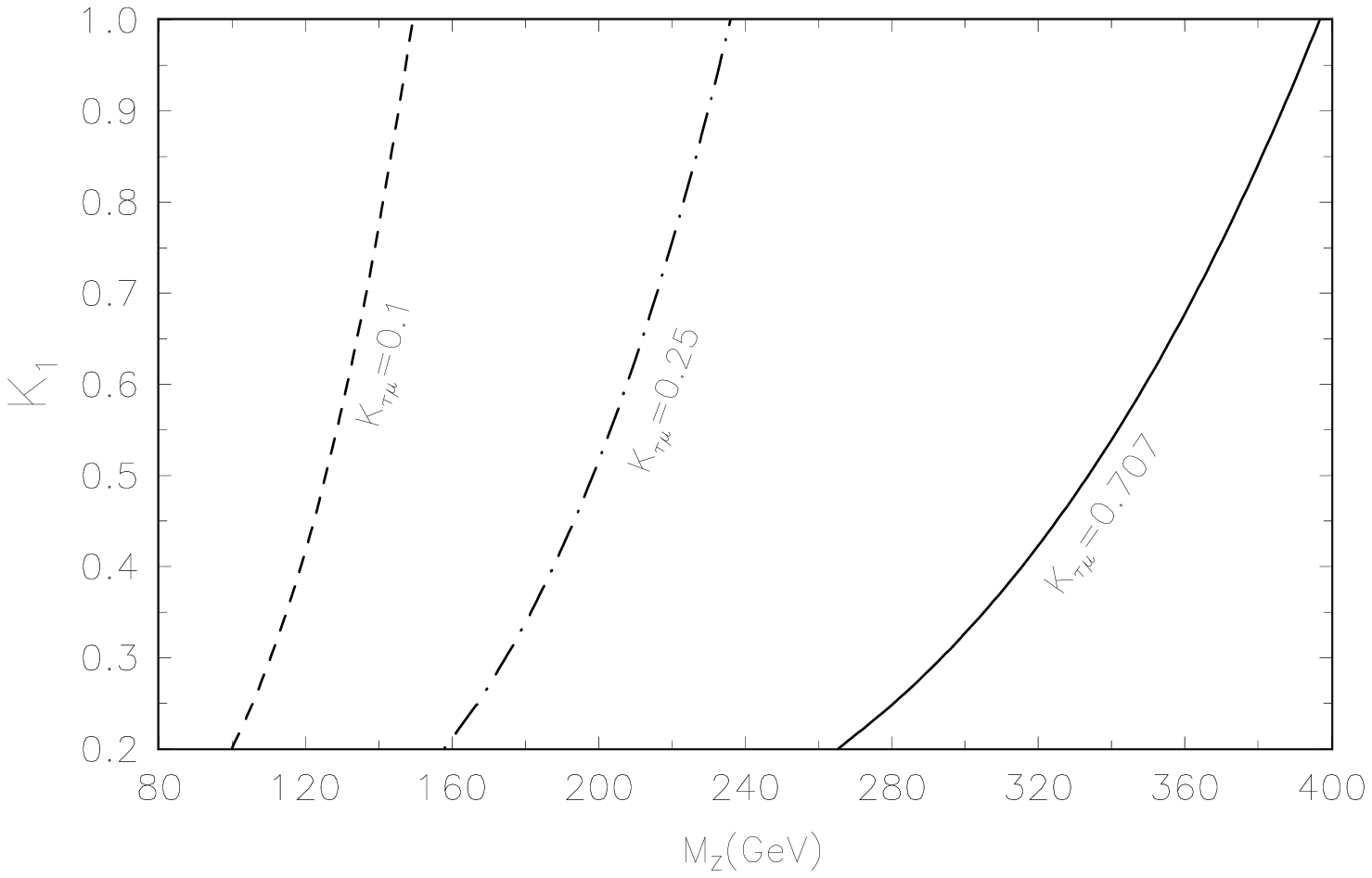}}\put(120,-10){Fig.1}
\end{picture}
\end{center}
\end{figure}
\begin{figure}[hb]
\begin{center}
\begin{picture}(250,200)(0,0)
\put(-50,0){\epsfxsize120mm\epsfbox{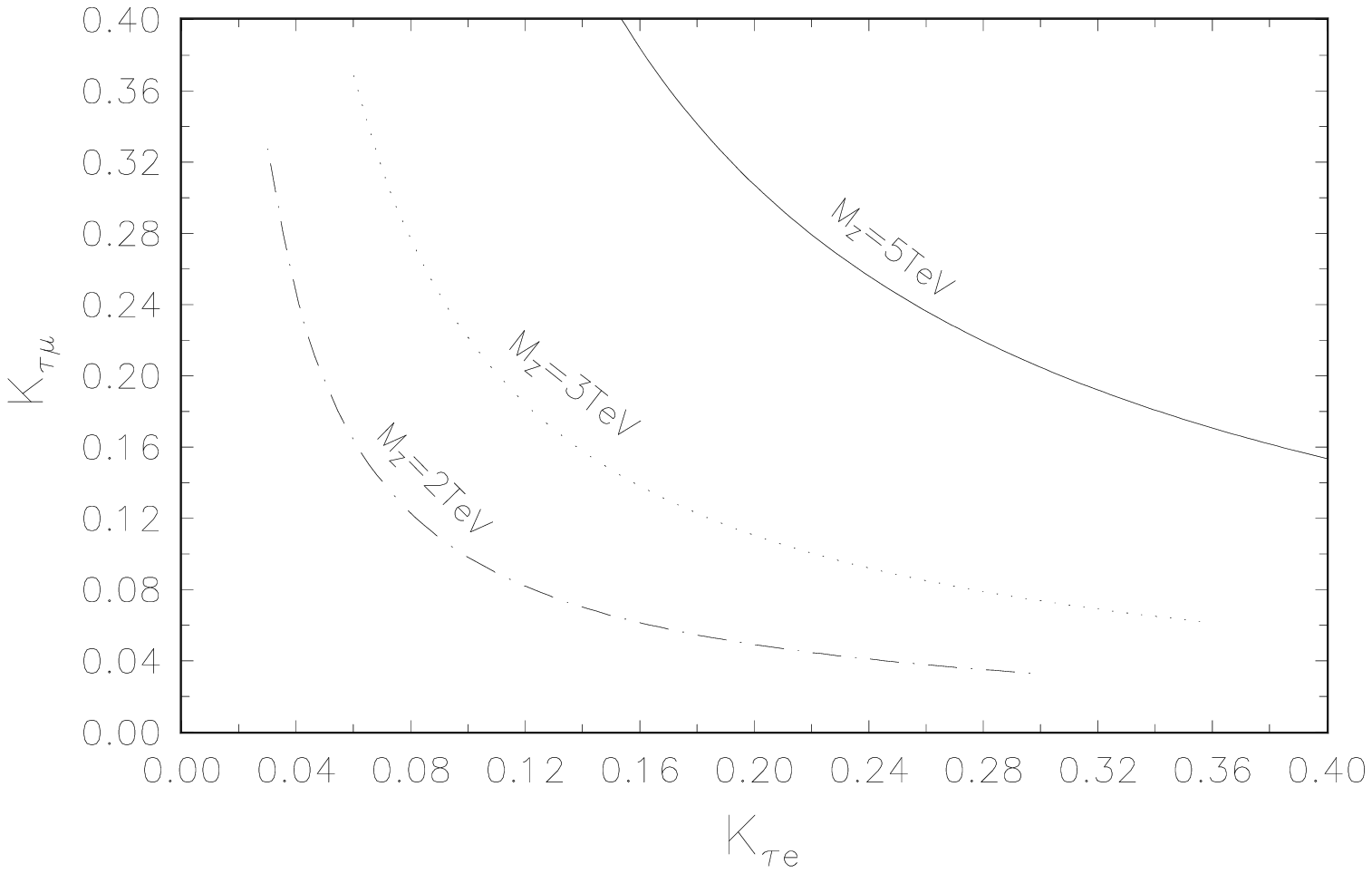}}\put(120,-10){Fig.2}
\end{picture}
\end{center}
\end{figure}
\newpage
\begin{figure}[pt]
\begin{center}
\begin{picture}(250,200)(0,0)
\put(-50,0){\epsfxsize120mm\epsfbox{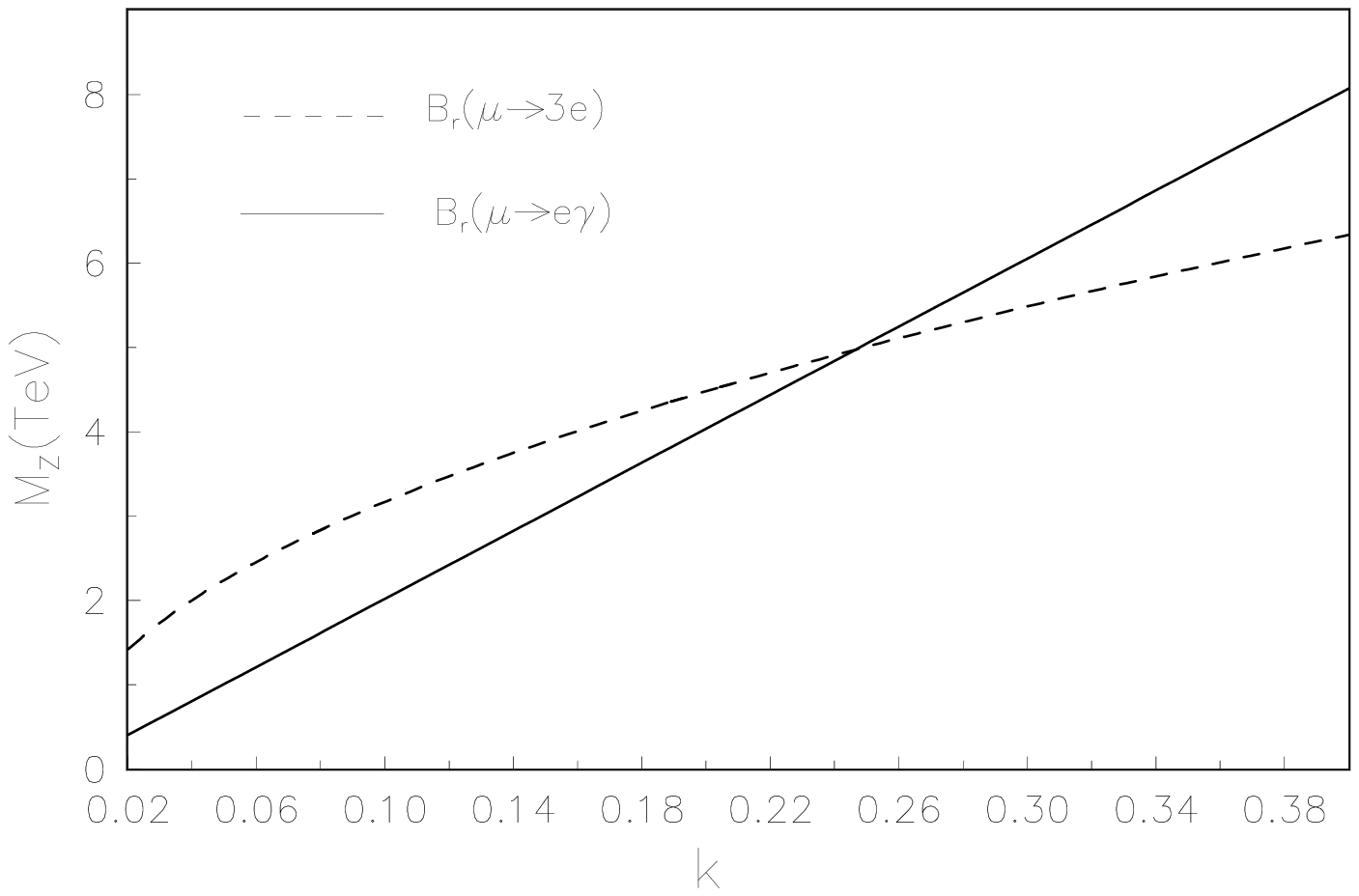}}\put(120,-10){Fig.3}
\end{picture}
\end{center}
\end{figure}
\end{document}